\documentstyle[prl,aps,twocolumn,epsf,floats]{revtex}  

\begin{document}

\title{The three-dimensional Ising model: A paradigm of liquid-vapor coexistence
in nuclear multifragmentation}

\author{C. M. Mader$^1$, A. Chappars$^2$}
\address{Hope College, Holland, MI 49423}

\author{J. B. Elliott, L. G. Moretto, L. Phair  and G. J. Wozniak}
\address{Nuclear Science Division, Lawrence Berkeley National Laboratory,
Berkeley, CA  94720}

\date{\today}
\maketitle

\begin{abstract}                
Clusters in the three-dimensional Ising model rigorously obey reducibility
and thermal scaling up to the critical temperature. The barriers extracted
from Arrhenius plots depend on the cluster size as $B \propto A^{\sigma}$
where $\sigma$ is a critical exponent relating the cluster size to the cluster
surface. All the Arrhenius plots collapse into a single Fisher-like scaling
function indicating liquid-vapor-like phase coexistence and the univariant
equilibrium between percolating clusters and finite clusters.  The compelling
similarity with nuclear multifragmentation is discussed.
\end{abstract}
\vspace{6pt}
Nuclear multifragmentation is a process occurring at the limits of nuclear
excitation, and, as such, portrays an appropriate richness and complexity.
While the fundamental problem of dynamics vs. statistics is still debated, it
appears ever more clearly that many thermal/statistical features underlie the
empirical body of data. In particular, two features associated with the
fragment multiplicities are found to be quite pervasive in all
multifragmentation reactions. They have been named ``reducibility" and
``thermal scaling''~\cite{ref:moretto,ref:beaulieu,ref:phair}.

Reducibility is the property of the $n$-fragment emission probability of
being expressible in terms of an elementary one-fragment emission
probability.  This property can occur only if fragments are created
independently from one another and it coincides with stochasticity.  Both
binomial, and its limiting form, Poissonian reducibilities have been
extensively documented
experimentally~\cite{ref:moretto,ref:beaulieu,ref:phair}.

Thermal scaling is the linear dependence of the logarithm of the one-fragment
probability with $1/T$ (an Arrhenius plot). It indicates that the emission
probability for fragment type $i$ has a Boltzmann dependence
    \begin{equation}
    \label{eqn:arrhen}
    p_i = p_0 e^{-B_i/T}
    \end{equation}
where $B_i$ is a barrier corresponding to the emission process.

The combination of these two empirical features powerfully attests to a
statistical mechanism of multifragmentation in general, and to liquid-vapor
coexistence specifically~\cite{ref:elliott2}.

Many statistical models have been proposed as an explanation for
multifragmentation. It is our intention here to identify a model which, on one
hand, is as simple as possible, yet on the other captures the essential
features observed in the experiments. Percolation in its many versions has
been widely used \cite{ref:stauffer1,ref:stauffer2,ref:bauer1,ref:bauer2}.
However, while being simple, it does not lend itself to a non-trivial thermal
study~\cite{ref:gujrati}. The three-dimensional Ising model satisfies both the
criteria of simplicity in its Hamiltonian and lends itself to a thermal
treatment with nontrivial results. While the Ising model has been widely
studied in terms of its continuous phase transition, the problem of
clusterization, of paramount interest to us here, has received relatively
little
attention~\cite{ref:stauffer3,ref:domb,ref:coniglio,ref:cambier,ref:kertesz,ref:wang1,ref:wang2,ref:demeo,ref:ferrenberg,ref:alonso}.
We will show that this model contains both features of reducibility and
thermal scaling. In showing the features of thermal scaling we will
demonstrate that the slopes of the Arrhenius plots associated with the
individual masses of the fragments, or the barriers in Eq.~(\ref{eqn:arrhen}),
portray a dependence on the fragment mass ($A$) of the form $B \propto
A^\sigma$, where $\sigma$ is a proper critical exponent which relates the mass
to the cluster surface. In addition, the individual Arrhenius plots for each
fragment mass can be absorbed into a single scaling function identical to that
of Fisher's droplet
model~\cite{ref:fisher1,ref:fisher2,ref:kiang,ref:stauffer4,ref:elliott1},
which defines the liquid-vapor coexistence line up to the critical
temperature.

The Hamiltonian of the Ising model has two terms: the interaction between
nearest neighbor ($n.n.$) spins in a fixed lattice and the interaction between
the fixed spins and an external applied field $H_{ext}$:
    \begin{equation}
    H = -J\sum_{i,j=\{n.n.\}} s_is_j  -  H_{ext}\sum_{i} s_i
    \label{eqn:ising}
    \end{equation}
where $J$ is the strength of the spin-spin interaction. In this model, the two
phases of the system are a liquid (clusters of sites with connected up spins)
and a gas (clusters of sites with connected down spins). In the absence of an
external field, the distinction between up and down spins vanishes. Thus in
the absence of an external field, these two phases are in coexistence below
the critical temperature, $T_c$. The critical temperature for the
three-dimensional Ising model has not been determined analytically; however,
Monte Carlo techniques have yielded a value of
$T_c=4.513~J/k_b$~\cite{ref:cambier}.

In the present study, Monte Carlo techniques are used to determine the equilibrium
cluster distribution as a function of temperature for a simple cubic
lattice with periodic boundary conditions. The lattice contains $50^3$ spins,
which is large but may not be large enough to represent an infinite system.  The
Swendsen-Wang algorithm~\cite{ref:wang2} was used to determine the equilibrium
spin configurations of the lattice for given temperatures, and clusters were
identified using the Coniglio-Klein~\cite{ref:coniglio} prescription.  Details about
the model can be found in reference~\cite{ref:longpaper}.

\begin{figure}[tb]
   \centerline{\epsfxsize=3.2in \epsffile{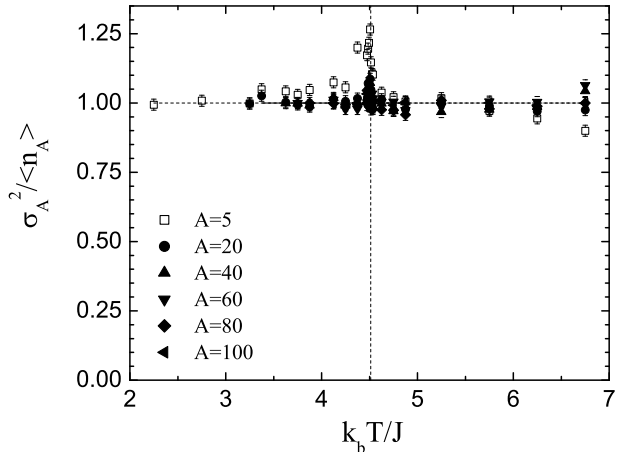}}
    \caption{Ratio of the variance to the mean for distributions of
    clusters with $A$ constituents. The dashed line corresponds to
    the critical temperature, $k_bT_c/J=4.513$.}
    \label{fig:varmean}
    \end{figure}

To test if the fragment distributions produced by the model exhibit binomial
or Poissonian characteristics, the ratio of the variance to the mean
($\sigma^2/\left<n_A\right>$) of the fragment distributions ($n_A$) for
various fragment sizes ($A$) as a function of temperature was determined. For
most clusters, the ratio is very close to one, which is the Poissonian limit,
over the entire temperature range considered here, which extends well beyond
the critical temperature. Only for the very lightest clusters ($A \lesssim
20$), does the ratio have a discontinuity at the critical temperature.
However, the ratio is near one even for these light clusters at all
temperatures away from the critical temperature. (See Fig.~\ref{fig:varmean}.)
Thus, for all but the smallest clusters, the distributions exhibit Poissonian
reducibility.
This signifies that the clusters are produced nearly independently of one
another, a feature also observed in percolation models, Fisher's droplet model
and nuclear fragmentation~\cite{ref:elliott2}.

If the fragment distributions also exhibit thermal reducibility, the
distributions must be of the form given in Eq.~(\ref{eqn:arrhen}). Thus in an
Arrhenius plot (a semi-log graph of the number of clusters of size $A$
($n_A$) vs. $1/T$), the distributions should be linear below the critical
temperature.

\begin{figure}[t]
   \centerline{\epsfxsize=3.4in \epsffile{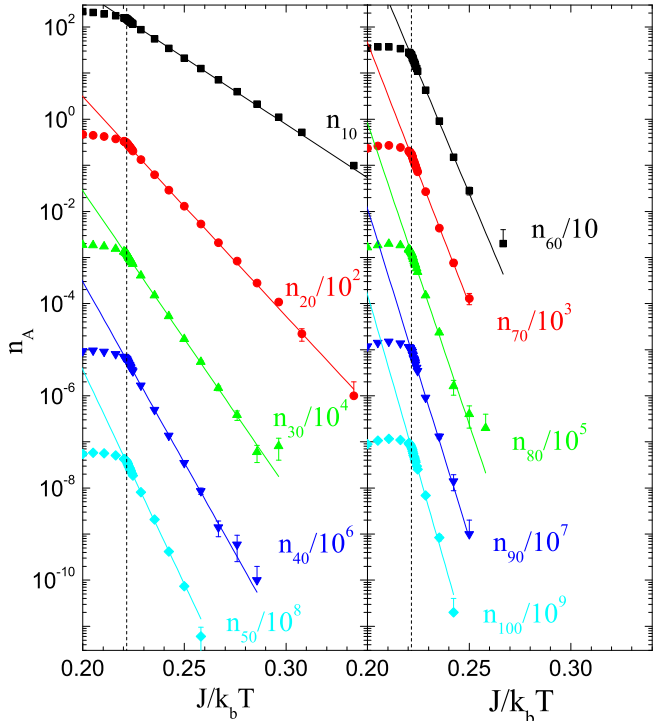}}
    \caption{Arrhenius plots of the cluster distributions. Statistical error bars are shown when
    they exceed the size of the data point. The lines are fits of the form given in
    Eq.~(\ref{eqn:arrhen}).  The critical temperature is
    indicated by the dashed line. }
    \label{fig:arrhen}
    \end{figure}

\begin{figure}[h]
   \centerline{\epsfxsize=2.7in \epsffile{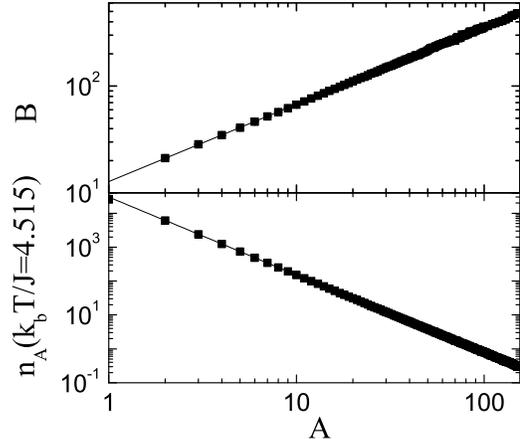}}
    \caption{The upper panel shows the extracted barriers from the fits to the
    cluster distributions. The line is a fit of the form given in
    Eq.~(\ref{eqn:bar}). The lower panel shows the power law behavior of the
    cluster distribution at $k_bT/J=4.515$. The line is a fit of the form given in
    Eq.~(\ref{eqn:tau}).   In both panels, error bars do not
    exceed the size of the data point.}
    \label{fig:barrier}
    \end{figure}

As shown in Fig.~\ref{fig:arrhen}, this is indeed the case over a wide range
of temperatures ($0<T<T_c$) and fragment sizes. While we have shown
distributions for clusters up to size $A=100$, the trend continues for larger
clusters, however statistics decrease significantly as the size of the cluster
increases. This linearity extends over more than four or five orders of
magnitude. It rigorously confirms the form of Eq.~(\ref{eqn:arrhen}) and
signifies the independent thermal formation of fragments controlled by a single
size-dependent barrier. This feature has been amply verified in nuclear
multifragmentation. By fitting the fragment distributions below the critical
temperature ($J/k_bT > 0.2216$), the barriers can be extracted. The
barriers for each cluster size are shown in Fig.~\ref{fig:barrier}.

These barriers should find their origin in the number of broken bonds
associated with a cluster. Therefore they should be well described by a
power-law:
    \begin{equation}
    B(A) = B_0 A^\sigma.
    \label{eqn:bar}
    \end{equation}
The fit of the extracted barriers according to Eq.~(\ref{eqn:bar}) is
$B=(12.77\pm0.04) A^{(0.7230\pm0.008)}$ and is remarkably good as shown in
Fig.~\ref{fig:barrier}. The value for $\sigma$ ($0.7230 \pm 0.0004$) is close
to 2/3, the value one would expect for spherical clusters of closely packed
spherical objects~\cite{ref:wang1}, and to 0.64 the value expected from
scaling relations~\cite{ref:domb}. Thus this is a new and effective method to
determine the critical exponent $\sigma$, never used before to our knowledge.

The features of reducibility and thermal scaling discussed above can be found
united in Fisher's formula for the cluster abundance in a vapor as a function
of cluster size and of temperature. The formula is
    \begin{equation}
    \label{eqn:fisher-arrh}
    n_A(T) = q_0 A^{-\tau} \exp (   \frac{A{\Delta}{\mu}}{T})
                               \exp (   \frac{c_0 A^{\sigma}}{T_c})
                               \exp ( - \frac{c_0 A^{\sigma}}{T})
    \end{equation}
where one can see the thermal scaling up to $T_c$ and the dependence of the
``barrier'' on the cluster mass through the critical exponent
$\sigma$~\cite{ref:fisher1,ref:fisher2}. It is interesting to explore
further the applicability of this formula to the Ising model.

In addition to the linear behavior of the Arrhenius plots below the critical
temperature, the Fisher droplet model also predicts that the cluster size
distribution at the critical point must follow a power law
    \begin{equation}
    n_A(T_c)=q_0 A^{-\tau}
    \label{eqn:tau}
    \end{equation}
where $q_0$ is fixed by the normalization relationship
    \begin{equation}
    q_0=\frac{\sum_{A=1}^\infty n_A(T_c)}{\sum_{A=1}^\infty A^{-\tau}}.
    \label{eqn:q0}
    \end{equation}
Away from the critical temperature, the cluster distribution should not follow
a pure power law. Thus, to determine $\tau$ without a prior knowledge of the
critical temperature, linear fits to the cluster distributions were determined
for all temperatures. At the critical temperature, the fit should have the
lowest $\chi^2/DoF$ and thus this fit should determine $\tau$. The results for
the best fit are shown in the lower panel of Fig.~\ref{fig:barrier}. From this
method, the critical temperature was found to be $(4.515\pm .010)J/k_B$ with a
best fit line of the form $n_A(T_c) = (30084\pm 7)A^{(-2.39671\pm 0.00017)}$.
This temperature is consistent with the value determined for infinite systems
($4.513J/k_B$) and the value of $\tau$ ($2.39671\pm 0.00017$) is close to the
expected value for an infinite system ($\tau^{\infty} = 2.21$). The fit also
determines $q_0$ which can be compared to the prediction of
Eq.~(\ref{eqn:q0}). The fit predicts $q_0^{fit} = 30084$ while $q_0^{\infty} =
27566$.  The power law fit of the fragment abundances at the critical
temperature is shown in Fig.~\ref{fig:barrier}.

\begin{figure}[h]
   \centerline{\epsfxsize=3.3in \epsffile{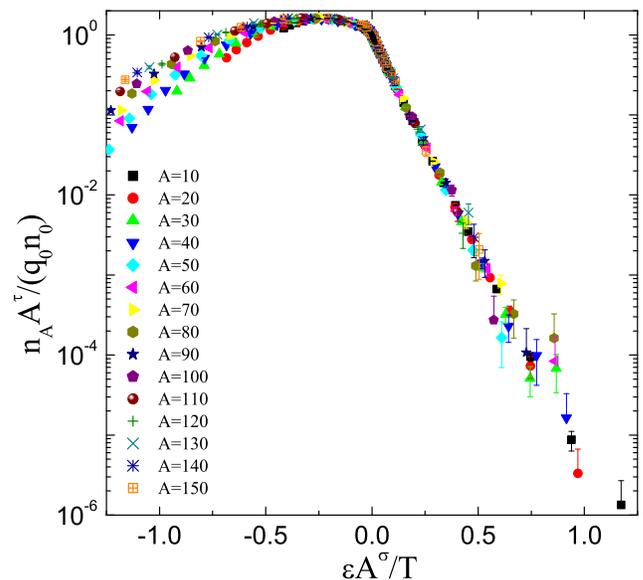}}
    \caption{Scaling behavior of cluster distributions.}
    \label{fig:scale}
    \end{figure}

    In the coexistence region, which in the Ising model should prevail for
$T<T_c$ when the magnetization is not constrained, the chemical potentials of
the liquid and gas phases are equal, thus Eq.~(\ref{eqn:fisher-arrh}) can be
rewritten as:
    \begin{equation}
    \label{eqn:scaled}
    n_A(T)A^{\tau}/q_0 = \exp( - c_0 A^\sigma \varepsilon /T )
    \end{equation}
with $\varepsilon = (T_c - T)/T_c$. Therefore, a graph of the scaled cluster
distributions ($n_A(T)A^{\tau}/q_0$) as a function of $\varepsilon A^\sigma/T$
should make the distributions of all cluster sizes collapse onto a single
curve. This scaling behavior can clearly be seen in Fig.~\ref{fig:scale}. This
nearly perfect collapse below the critical temperature extends over six orders
of magnitude for the broadest range of cluster sizes and it is perfectly
linear. Therefore the three-dimensional Ising model and fluids belong to the
same class of universality and can be described by Fisher's droplet model. The
Ising clusters constructed here can be properly thought of as ``vapor'' in
equilibrium with the ``liquid'' percolating cluster. Coexistence of the two
phases is determined by the observation that the empirical scaling requires
$\Delta \mu = 0$. The fact that both the three-dimensional Ising model and the
experimental nuclear multifragmentation data obey the same scaling predicted by
Fisher's droplet model indicates that nuclear multifragmentation can indeed be
identified as the clustering (non-ideality) in a nuclear vapor in equilibrium with the nuclear
liquid~\cite{ref:elliott2}.

We note that the value of $\sigma$ determined in this work is larger than
$0.64$, the usual value of $\sigma$ from studies of the thermal properties of
three-dimensional Ising systems. There are several possible explanations for
the differences between the value of $\sigma$ determined in this study and
the values found from other studies. First, while statistical errors are
small, systematic errors may not be so small. Finite size effects of the
$50^3$ lattice could lead to an overabundance of small clusters. This is
caused by the fact that large clusters near the edge of the lattice are unable
to grow beyond the lattice edges easily, even though periodic boundary
conditions are used in this model. The effects of lattice size as well as
other systematic errors are discussed in greater detail
in~\cite{ref:longpaper}. An over-abundance of small clusters, with a
corresponding suppression of large clusters, would increase the slope of the
cluster distribution leading to larger than expected values of $\tau$ and
$q_0$, just as we have observed.

    In conclusion, we have shown that the clusterization in the Ising model, like
nuclear multifragmentation, portrays reducibility and thermal scaling.  In
addition, the Arrhenius plots allow for the extraction of ``barriers'' which
are found to have a dependance of $B\propto A^\sigma$, where $\sigma$ (which is
close to 2/3) is a critical exponent.  The reducibility and thermal scaling
features in the Ising model can be incorporated into a Fisher-like scaling
with $\Delta \mu=0$, which is obeyed rigorously over the explored temperature
range below the critical temperature.  Thus the observed clusters can be
interpreted as the non-idealities of a vapor in equilibrium with a liquid.
Finally, nuclear multifragmentation, which is seen to share all the scaling
observed here, should be similarly interpreted as the clusterization of a
nuclear vapor in equilibrium with its liquid.

\vfill
\noindent {\bf Acknowledgements}

This work was supported by the National Science Foundation under grants
NSF-RUI 9800747 and NSF-REU 9876955 and the Nuclear Physics Division of the
US Department of Energy under contract DE-AC03-76SF00098. One of us (CMM)
acknowledges support from LBNL during her sabbatical visit.

\vfill
\noindent {\bf Present Addresses}

\noindent
$^1$ Nuclear Science Division, Lawrence Berkeley National Laboratory,
Berkeley, CA\ \ 94720\\
$^2$ Marietta College, Marietta, OH\ \ 45750

\end{document}